\documentclass[prl,twocolumn,superscriptaddress,floatfix,showpacs,amsmath,longbibliography,nofootinbib,amssymb]{revtex4-1}
\usepackage[colorlinks=true,citecolor=blue,linkcolor=blue]{hyperref}
\usepackage[usenames]{color}
\usepackage{subfigure}
\usepackage{dcolumn}
\usepackage{mathrsfs}% Align table columns on decimal point
\usepackage{bm}% bold math
\usepackage{amsmath,amssymb,epsfig,float,graphics}
\usepackage{appendix}
\newcommand{\bk}{\mathbf{k}}
\newcommand{\br}{\mathbf{r}}
\newcommand{\bR}{\mathbf{R}}
\newcommand{\cA}{\mathcal{A}}
\newcommand{\cH}{\mathcal{H}}
\DeclareMathOperator{\erf}{erf}

\begin{document}
\baselineskip=0.45 cm

\title{Probing Lorentz-Invariance-Violation Induced Nonthermal Unruh Effect in Quasi-Two-Dimensional Dipolar Condensates}

\author{Zehua Tian}
\email{tianzh@ustc.edu.cn}
\affiliation{CAS Key Laboratory of Microscale Magnetic Resonance and School of Physical Sciences, University of Science and Technology of China, Hefei 230026, China}
\affiliation{CAS Center for Excellence in Quantum Information and Quantum Physics, University of Science and Technology of China, Hefei 230026, China}

\author{Longhao Wu}
\affiliation{CAS Key Laboratory of Microscale Magnetic Resonance and School of Physical Sciences, University of Science and Technology of China, Hefei 230026, China}
\affiliation{CAS Center for Excellence in Quantum Information and Quantum Physics, University of Science and Technology of China, Hefei 230026, China}

\author{Liang Zhang}
\affiliation{CAS Key Laboratory of Microscale Magnetic Resonance and School of Physical Sciences, University of Science and Technology of China, Hefei 230026, China}
\affiliation{CAS Center for Excellence in Quantum Information and Quantum Physics, University of Science and Technology of China, Hefei 230026, China}

\author{Jiliang Jing}
\affiliation{Department of Physics, Key Laboratory of Low Dimensional Quantum Structures and Quantum Control of Ministry of Education, and Synergetic Innovation Center for Quantum Effects and Applications, Hunan Normal University, Changsha, Hunan 410081, P. R. China}

\author{Jiangfeng Du}
\email{djf@ustc.edu.cn}
\affiliation{CAS Key Laboratory of Microscale Magnetic Resonance and School of Physical Sciences, University of Science and Technology of China, Hefei 230026, China}
\affiliation{CAS Center for Excellence in Quantum Information and Quantum Physics, University of Science and Technology of China, Hefei 230026, China}
\affiliation{Hefei National Laboratory, University of Science and Technology of China, Hefei 230088, China}

\begin{abstract}
The Unruh effect states an accelerated particle detector registers a thermal response when moving through the Minkowski vacuum, and its
thermal feature is believed to be inseparable from Lorentz symmetry: Without the latter, the former disappears.
Here we propose to observe analogue circular Unruh effect
using an impurity atom in a quasi-two-dimensional Bose-Einstein condensate (BEC) with dominant 
dipole-dipole interactions between atoms or molecules in the ultracold gas. Quantum fluctuations in the condensate possess a Bogoliubov spectrum 
$\omega_\bk=c_0|\bk|f(\hbar\,c_0|\bk|/M_\ast)$, working as an analogue Lorentz-violating quantum field with the Lorentz-breaking scale $M_\ast$, 
and the impurity acts as an effective Unruh-DeWitt detector thereof. When the detector travels close to the sound speed, 
observation of the Unruh effect in our quantum fluid platform becomes experimentally 
feasible. In particular, the deviation of the Bogoliubov spectrum from the Lorentz-invariant case is highly engineerable through the relative strength of the dipolar and contact interactions, and thus a viable laboratory tool is furnished to experimentally investigate whether the thermal characteristic of Unruh effect is robust to the breaking of
Lorentz symmetry.

\end{abstract}

%\pacs{85.25.Dq, 84.40.Az, 04.80.Cc, 04.62.+v, 98.80.Cq}
\baselineskip=0.45 cm
\maketitle
\newpage

%%%%%%%%%%%%%%%%%%%%%%%%%%%%%%%%%%%%%%%%%%%%%%%%%%%%%%%%%%%%
\textit{Introduction.---} One of the surprising fundamental consequences of relativistic quantum field theory is that 
the concept of particle number is observer dependent. A prominent paradigm is the so-called Unruh effect \cite{PhysRevD.14.870}: 
In the view of an uniformly accelerating observer, the Fock vacuum state of quantum field in the Minkowski spacetime appears as  
a thermal state rather than a zero-particle state. The corresponding characteristic temperature is  proportional to the observer's 
acceleration $a$, given by $k_\text{B}T_\text{U}=\frac{\hbar\,a}{2\pi\,c}$.
To produce a measurable temperature for fundamental quantum fields, extremely huge accelerations are required (e.g., smaller than $1\,\mathrm{Kelvin}$ even for accelerations as high as $10^{20}\,\mathrm{m/s^2}$), and thus until now,
the direct experimental confirmation of the Unruh effect still remains elusive.

Analogue gravity  \cite{PhysRevLett.46.1351, Analogue-Gravity} opens up a new route  
to study various phenomenas predicted by relativistic quantum field theory, e.g., Hawking effect \cite{PhysRevLett.85.4643, PhysRevLett.103.087004, PhysRevLett.105.240401, PhysRevLett.85.4643, Jeff3, PhysRevLett.106.021302, PhysRevLett.104.250403, PhysRevLett.117.121301, Jeff1,  PhysRevLett.118.061301, Tian, Jeff4, Jeff2, PhysRevLett.122.010404, PhysRevLett.123.161302, PhysRevResearch.2.043285, PhysRevResearch.2.023107, PhysRevD.105.124066, Tian-H}, cosmological particle production \cite{PhysRevA.69.033602, doi:10.1126/science.1237557, PhysRevD.82.105018, PhysRevLett.91.240407, PhysRevLett.94.220401, PhysRevD.95.125003, PhysRevLett.99.201301, PhysRevX.8.021021,PhysRevD.100.065003,  PhysRevLett.118.130404, PhysRevLett.123.180502, PhysRevA.103.023322, PhysRevLett.128.090401}, and dynamical Casimir effect \cite{PhysRevLett.103.147003, PhysRevB.84.174521, DCE, PhysRevLett.109.220401, ki4234, PhysRevX.8.011031, PhysRevLett.112.036406, physics2010007, PhysRevLett.124.140503}, in a variety of electronic, acoustic, optical and even magnetic and superconducting settings.
Recently, analogue gravity program for observing the Unruh effect has been successfully theoretically put forward \cite{PhysRevLett.83.256, PhysRevLett.125.241301,  PhysRevLett.100.091301, PhysRevLett.101.110402, PhysRevResearch.2.042009, PhysRevLett.125.213603, PhysRevA.103.013301, PhysRevLett.123.156802, PhysRevLett.126.117401, PhysRevLett.118.161102, PhysRevD.101.065015, PhysRevLett.122.053604, PhysRevA.102.033506}, and through a BEC system Hu \emph{et al.} experimentally realized the analogue Unruh effect relied on functional equivalence (i.e., simulating two-mode squeezed mechanics) \cite{Unruh-effect1}. Furthermore, in the quantum field theory, 
the contributions from the trans-Planckian modes as seen by an inertial observer are indispensable for deriving the Unruh effect with 
Bogoliubov transformation method. This particular feature makes the Unruh effect a 
potentially important arena for understanding and exploring implications of 
trans-Planckian physics \cite{NICOLINI2011303, PhysRevD.77.124032, PhysRevD.92.024018, Kajuri_2016, Hossain_2016, PhysRevD.94.104055, PhysRevLett.123.041601, PhysRevD.103.085010, PhysRevD.97.025008}, and even the detecting means to probe some candidate theories of quantum gravity that may modify the trans-Planckian modes significantly. 
This modification usually accompanied by the breaking of Lorentz symmetry may even challenge the equivalence between Unruh effect and Hawking effect since the latter 
appears to be robust to high energy modifications of the dispersion relation \cite{PhysRevD.51.2827} while the former is not so immune and will lose its conventional
thermal interpretation \cite{NICOLINI2011303, PhysRevD.77.124032, PhysRevD.92.024018, Hossain_2016, PhysRevLett.123.041601} (see more details in
the following). It is thus of great interest to experimentally investigate the consequences of trans-Planckian physics in a microscopically well-understood setup in a regime, that is inaccessible for quantum fields in real relativistic scenarios.

In this paper, we propose to study the interplay between the Unruh effect and trans-Planckian physics with an experimentally accessible platform consisting of a dipolar BEC \cite{BARANOV200871} and an immersed impurity \cite{PhysRevLett.94.040404, PhysRevLett.91.240407}.
From the perspective of analog, the density fluctuations in the condensate possessing
trans-Planckian spectra leading to strong departures from Lorentz invariance \cite{PhysRevA.73.031602, PhysRevLett.118.130404, PhysRevA.97.063611},
resemble  Lorentz-violating quantum field (LVQF), while the impurity, analogously dipole coupled to the density fluctuations, 
is modeled as an Unruh-DeWitt detector coupled to the LVQF. We will show that for the Lorentz-invariant (LI) spectrum the Unruh-DeWitt detector for an accelerated circular path indeed experiences a similar thermal response, while yields significant changes of this standard thermal feature when the spectra strongly deviate from Lorentz invariance. As far as we know, this represents the first example within analogue gravity where Unruh effect without thermality caused by 
the breaking of Lorentz symmetry can become experimentally manifest.

%%%%%%%%%%%%%%%%%%%%%%%%%%%%%%%%%%%%%%%%%%%%%%%%%%%%%%%%%%%%

\begin{figure}
\centering
\includegraphics[width=0.38\textwidth]{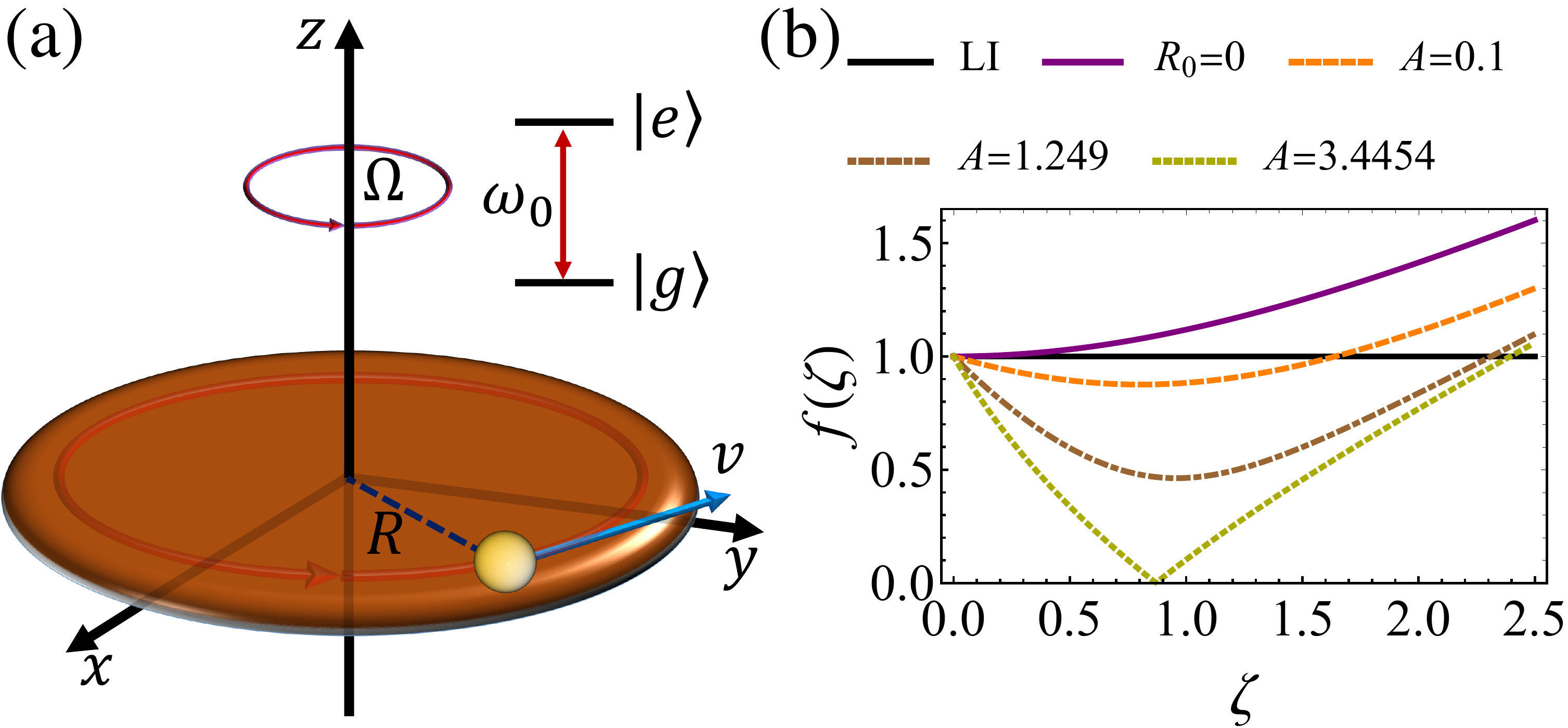}
\caption{(a) A two-level atom immersed in a quasi-2D dipolar BEC, moving along a circular trajectory with the radius of the orbit $R$, an angular velocity $\Omega$ and the corresponding linear speed $v$. (b) The dimensionless function $f$ shown in \eqref{Dispersion} as a function of $\zeta=\hbar\,c_0k/M_\ast$.
For the LI case, $f=1$. $R_0=0$ denotes the contact interaction case, where $f$ is independent of $A$. 
For DDI dominance, $R_0=\sqrt{\pi/2}$, $f$ dips below 1 for an interval of $\zeta$. Note that 
$f$ becomes negative when $A>A_c=3.4454$, which means the spectrum of quasiparticle 
becomes unstable.}\label{fig1}
\end{figure}

%%%%%%%%%%%%%%%%%%%%%%%%%%%%%%%%%%%%%%%%%%%%%%%%%%%%%%%%%%%%
\textit{Lorentz-violating quantum field and analogue Unruh-DeWitt detector in dipolar BEC.---} 
As schematically shown in Fig. \ref{fig1}, we establish the connection between an impurity immersed in a quasi-two-dimensional (quasi-2D) dipolar BEC with the Unruh-DeWitt detector model \cite{PhysRevD.14.870, birrell1984quantum}, inspired by the seminal atomic quantum dot idea introduced in Refs. \cite{PhysRevLett.94.040404, PhysRevLett.91.240407, PhysRevD.69.064021}.

We begin with the Lagrangian density of an interacting Bose gas comprising atoms or molecules of mass $m$, 
\begin{eqnarray}\label{Lagrangian}
\nonumber
\mathcal{L}&=&\frac{i\hbar}{2}(\Psi^\ast\partial_t\Psi-\partial_t\Psi^\ast\Psi)-\frac{\hbar^2}{2m}|\nabla\Psi|^2-V_\text{ext}|\Psi|^2
\\
&&-\frac{1}{2}|\Psi|^2\int\,d^3\bR^\prime\,V_\text{int}(\bR-\bR^\prime)|\Psi(\bR^\prime)|^2,
\end{eqnarray}
where $\bR=(\br, z)$ are spatial 3D coordinates. The interaction reads $V_\text{int}(\bR-\bR^\prime)=g_c\delta^3(\bR-\bR^\prime)+3g_d\{[1-3(z-z^\prime)^2/|\bR-\bR^\prime|^2]/|\bR-\bR^\prime|^3\}/4\pi$, where $g_c=4\pi\hbar^2a_c/m_B$ represents the contact interaction strength, with $a_c$ being the $s$-wave scattering length; the dipole-dipole interaction (DDI) strength $g_d=\mu_0\mu_m^2/3$, with $\mu_0$ and $\mu_m$ being the permeability of vacuum and the magnetic dipole moment that is polarized to the $z$ direction, respectively. Moreover, the gas is trapped by an external potential, $V_\text{ext}(\bR)=m_B\omega^2\br^2/2+m_B\omega^2_zz^2/2$, and is strongly confined along the $z$ axis, with aspect ratio 
$\kappa=\omega_z/\omega\gg1$ over the whole time evolution. As a result of that, the motion of the Bose gas along the $z$ axis is frozen to the ground state with a Gaussian form, $\rho_z(z)=(\pi\,d^2_z)^{-1/2}\exp[-z^2/d^2_z]$, with $d_z=\sqrt{\hbar/m_B\omega_z}$. Therefore, the whole system effectively reduces to a quasi-2D one 
which can ensure stability in the DDI-dominated regime \cite{PhysRevA.73.031602}. Finally, we may assume that the Bose gas is condensed to the zero-momentum state 
with an area density $\rho_0$. 

Within the Bogoliubov theory of small excitations on top of the condensate \cite{pethick2008bose, pitaevskiui2016bose}, density fluctuations in Heisenberg representation can be written as
 $\delta\hat{\rho}(t,\br)=\sqrt{\rho_0}\int[d^2\bk/(2\pi)^2](u_\bk+v_\bk)[\hat{b}_\bk(t)e^{i\bk\cdot\br}+\hat{b}^\dagger_\bk(t)e^{-i\bk\cdot\br}]$ which closely resemble the quantum field in terms of bosonic operators $\hat{b}_\bk(t)=\hat{b}_\bk\,e^{-i\omega_\bk\,t}$, satisfying the usual Bose commutation rules 
$[\hat{b}_\bk, \hat{b}_{\bk^\prime}^\dagger]=(2\pi)^2\delta^2(\bk-\bk^\prime)$. Note that
$u_\bk=(\sqrt{\cH_\bk}+\sqrt{\cH_\bk+2\cA_\bk})/2(\cH^2_\bk+2\cH_\bk\cA_\bk)^{1/4}$ and $v_\bk=(\sqrt{\cH_\bk}-\sqrt{\cH_\bk+2\cA_\bk})/2(\cH^2_\bk+2\cH_\bk\cA_\bk)^{1/4}$ are Bogoliubov parameters, and the quasiparticle frequency 
$\omega_\bk=\sqrt{\cH^2_\bk+2\cH_\bk\cA_\bk}$ with 
$\cH_\bk=k^2/2m$ and $\cA_\bk=\rho_0V^\text{2D}_\text{int, 0}(k)$ \cite{PhysRevA.97.063611}. Here $k=|\bk|$ and the Fourier transformation of DDI $V^\text{2D}_\text{int, 0}(k)=g^\text{eff}_0(1-\frac{3R_0}{2}kd_zw[\frac{kd_z}{\sqrt{2}}])$, with $w[x]=\exp[x^2](1-\erf[x])$,
an effective contact coupling $g^\text{eff}_0=\frac{1}{\sqrt{2\pi}d_z}(g_c+2g_d)$, and the dimensionless ratio $R_0=\sqrt{\pi/2}/(1+g_c/2g_d)$.
The parameter $R_0$ could be tunable via Feshbach resonance \cite{PhysRevLett.81.69, Inouye1998Observation, RevModPhys.82.1225, TIMMERMANS1999199} and rotating polarizing field \cite{PhysRevLett.89.130401, PhysRevLett.120.230401}, ranging from $R_0=0$ (when $g_d/g_c\rightarrow0$, i.e., contact dominance), to $R_0=\sqrt{\pi/2}$ (when $g_d/g_c\rightarrow\infty$, i.e., DDI dominance).

The density fluctuations described above closely resemble a LVQF with a 
explicit dispersion relation given by 
\begin{eqnarray}\label{Dispersion}
\omega_\bk=c_0k\sqrt{1-\frac{3R_0}{2}\sqrt{A}\zeta\,w\bigg[\sqrt{\frac{A}{2}}\zeta\bigg]
+\frac{\zeta^2}{4}}
=c_0kf(\zeta),
\end{eqnarray}
where $\zeta=\hbar\,c_0k/M_\ast$, $c_0=\sqrt{g^\text{eff}_0\rho_0/m_B}$ is the speed of sound, 
$A=g^\text{eff}_0\rho_0/\hbar\omega_z$ represents the effective chemical potential as measured relative to the transverse trapping,  and $M_\ast=m_Bc^2_0$ is the analog energy scale 
of Lorentz violation. This dispersion relation \eqref{Dispersion} is approximately 
Lorentz invariant $(f(\zeta)\simeq1)$ for $\zeta\ll1$. 
By appropriately setting the relevant parameters $A$ and $R_0$, 
the dispersion could be analogously superluminal $(f(\zeta)>1)$ and subluminal 
$(f(\zeta)<1)$. In Fig. \ref{fig1}, we plot the function $f(\zeta)$ shown in \eqref{Dispersion}
to see how the Lorentz invariance  is violated in this dispersion.
For the DDI dominance, $R_0=\sqrt{\pi/2}$, the analogous subluminal spectrum develops a roton minimum 
for sufficiently large $A$, and the Lorentz invariance is strongly broken near $\zeta\simeq0.9$ \cite{PhysRevA.97.063611}.

In order to probe the analogue LVQF in the dipolar BEC, we use an impurity as the analogue Unruh-DeWitt detector, which   
consists of a two-level atom ($1$ and $2$) and its motion is supposed to be externally imposed by a tightly confining and relatively moving trap potential, so that its only degrees of freedom are the internal ones. Furthermore, the impurity is assumed to be controlled by a driving of a monochromatic external electromagnetic field 
at the frequency $\omega_L$ close to resonance with $1\rightarrow2$ transition $\omega\simeq\omega_{21}$, with a Rabi frequency $\omega_0$. Then the Hamiltonian of the whole system can be written as 
\begin{eqnarray}\label{System-Hamiltonian}
\nonumber
H(t)&=&\int\,d^2\bk\hbar\omega_{\bf k}\hat{b}^\dagger_{\bf k}\hat{b}_{\bf k}+\hbar\omega_{21}|2\rangle\langle2|-\bigg(\frac{\hbar\omega_0}{2}e^{-i\omega_Lt}|2\rangle\langle1|
\\
&&+\mathrm{H.c.}\bigg)
+\sum_{s=1, 2}g_s\hat{\rho}(\br_A(t))|s\rangle\langle\,s|,
\end{eqnarray}
where the last term denotes the collisional coupling between the impurity and Bose gas, and $\hat{\rho}(\br_A)=\hat{\psi}^\dagger(\br_A)\hat{\psi}(\br_A)\simeq\rho_0+\delta\hat{\rho}(\br_A)$ represents the field density operator of the Bose gas with $\br_A(t)$ being the time-dependent position of the impurity.

In the rotated $|g, e\rangle=(|1\rangle\pm2\rangle)/\sqrt{2}$ basis, the Rabi frequency $\omega_0$ determines the splitting between the $|g, e\rangle$ states, 
while the detuning $\delta=\omega_L-\omega_{21}$ gives a coupling terms. In such case, the impurity immersed in the condensate is 
collisionally coupled to the Bose gas via two channels \cite{PhysRevLett.118.045301, PhysRevResearch.2.042009, PhysRevD.103.085014}:
The first term resembles the interaction of a static charge to an external scalar potential and can be canceled through proper tuning of 
the interaction constants $g_{1, 2}$ (e.g., via Feshbach resonance  \cite{PhysRevLett.81.69, Inouye1998Observation, RevModPhys.82.1225, TIMMERMANS1999199}), while the second term resembles a standard electric-dipole coupling. Choosing proper detuning $\delta$ to exactly compensate the coupling to the average density, we can finally find the 
impurity-fluctuations interaction Hamiltonian 
\begin{eqnarray}\label{interaction2}
H_\text{int}=g_-(e^{i\omega_0\tau}\sigma^+e^{-i\omega_0\tau}\sigma_-)\delta\hat{\rho}({\bf r}_A(\tau), t_A(\tau)),
\end{eqnarray}
reproducing the usual Unruh-DeWitt detector-field interaction  with $g_-=(g_1-g_2)/2$ satisfying $\hbar\delta/2+g_-\rho_0=0$. However, here the LVQF is coupled to 
the detector.

Note that when $\zeta=\hbar\,c_0k/M_\ast\ll1$ the density fluctuations resemble massless scalar field with spectrum, $\omega_\bk\simeq\,c_0k$, the linearly moving impurity remains unexcited when its velocity satisfying $v<c_0$ , while behaves dramatically differently when moving at a supersonic speed $v\gtrsim\,c_0$.
Specifically, although the ``charge neutrality" of the impurity rules out Bogoliubov-Cherenkov emission \cite{PhysRevLett.97.260403, PhysRevA.70.013608}, 
the anomalous Doppler effect may induce it to be excited from its ground state while emitting Bogoliubov phonon and still conserving energy.
This is the analogue Ginzburg emission for superluminal moving particles \cite{1986ZhPmR, 1996}, occurring in BEC.
When the spectrum breaks the Lorentz symmetry satisfying $\omega_{\bk}=c_0kf(\zeta)$: If $0<f(\zeta)<1$ for an interval of $\zeta$,
the impurity would get excited when its velocity exceeds the critical $v_c=c_0f_c(\zeta)$ with $f_c(\zeta)=\inf\,f(\zeta)$ \cite{PhysRevD.103.085014}.
This particular property provides us a potential effective tool to constrain on the possible families of modified dispersion relations with the experimental 
results of the Relativistic Heavy Ion Collider \cite{PhysRevLett.116.061301}. We will in the following consider a circularly moving impurity to observer the circular Unruh 
effect \cite{BELL1983131, BELL1987488,UNRUH1998163}, and in particular examine whether the circular Unruh effect is robust to the breaking of Lorentz symmetry.

%%%%%%%%%%%%%%%%%%%%%%%%%%%%%%%%%%%%%%%%%%%%%%%%%%%%%%%%%%%%%
\begin{figure*}
\centering
\includegraphics[width=0.66\textwidth]{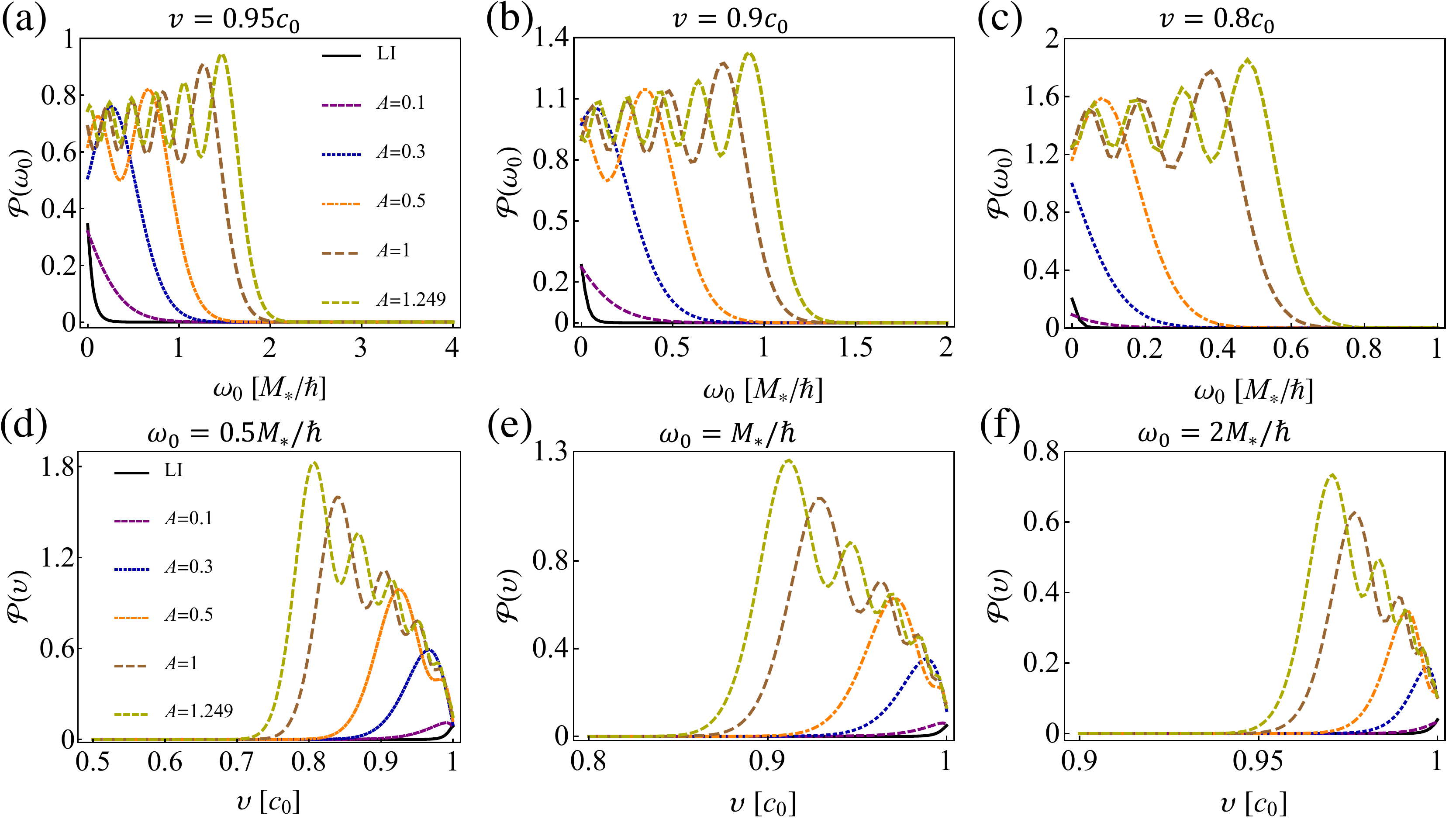}
\caption{Transition rates shown in \eqref{MFTR} for different effective chemical potential $A$, in units of $\frac{g^2_0\rho_0m_B}{2\hbar^3}$: (a), (b), (c) as a function of 
the detector's energy spacing $\omega_0$ with fixed velocity; (d), (e), (f) as a function of the velocity $v$ with fixed energy spacing of the detector. Here we assume that the DDI dominance case ($R=\sqrt{\pi/2}$) is vailid.}\label{fig2}
\end{figure*}

%%%%%%%%%%%%%%%%%%%%%%%%%%%%%%%%%%%%%%%%%%%%%%%%%%%%%%%%%%%
\textit{Spontaneous excitation of the circularly moving detector.---} 
If the detector moves with a circular trajectory $(c_0t(\tau), {\bf x}(\tau))=(c_0\gamma\tau, R\cos(\Omega\gamma\tau), R\sin(\Omega\gamma\tau), 0)$, with constant radius $R$, the angular velocity $\Omega$, the usual relativistic factor $\gamma=1/\sqrt{1-R^2\Omega^2/c^2_0}$, and the corresponding acceleration $a=\Omega^2\gamma^2R$,
we find the transition rate of the detector from its ground state to excited state \cite{Supplemental-Material}
\begin{eqnarray}\label{MFTR}
\nonumber
{\cal P}(\omega_0)&=&\frac{g^2_-\rho_0m_B}{2\hbar^3}\int^\infty_0d\zeta\frac{\zeta^2}{\gamma\,f(\zeta)}\sum^{\infty}_{m=-\infty}J^2_m(\tilde{M}\zeta)
\\      
&\times&\delta\bigg(\zeta\,f(\zeta)-\frac{1}{\tilde{M}}\bigg(mv-\frac{\tilde{E}}{\gamma}\bigg)\bigg),
\end{eqnarray}
where $\tilde{M}=RM_\ast/\hbar\,c_0$, $\tilde{E}=R\omega_0/c_0$, and $v=R\Omega/c_0$ are dimensionless parameters.
Note that we focus on the detector's speed in the preferred Lorentz frame satisfying $0\le\,v<1$.

For the LI scenario where $R_0=0$ and $\omega_\bk=c_0k\sqrt{1+\zeta^2/4}\approx\,c_0k$ for $k\ll\,k_c=M_\ast/c_0\hbar$, we find in the ultrarelativistic limit, $\gamma\gg1$, the equilibrium population of the upper level relative to the lower is \cite{Supplemental-Material}
\begin{eqnarray}
\frac{{\cal P}(\omega_0)}{{\cal P}(-\omega_0)}=\frac{12c^2_0\omega^2_0}{5a^2}\exp\big(-\sqrt{\frac{24}{5}}\frac{c_0\omega_0}{a}\big),
\end{eqnarray} 
when the energy splitting of the detector is not too small $\omega_0\gg\,a/c_0$ \cite{BELL1983131}, which leads to an effective temperature 
\begin{eqnarray}
T_\text{eff}=\frac{\sqrt{5}\hbar\,a}{2\sqrt{6}k_Bc_0}.
\end{eqnarray}
This temperature is higher by a factor $\sqrt{5/6}\pi$ than the Unruh temperature for the linear acceleration, and higher by a factor $\sqrt{5/2}$ than 
the Unruh temperature for the real massless scalar field case as a result of the Bogoliubov transformation for the quasiparticles.

If $(\zeta\,f(\zeta))^\prime>0$, we can find in the limit, $\tilde{M}\rightarrow\infty$, the transition rate \eqref{MFTR} behaves quite differently for $0<v<f_c$ and $f_c<v<1$, where $f_c=f(\zeta_c)$ is a global minimum at $\zeta=\zeta_c>0$. Specifically, for the former, the transition rate ${\cal P}(\omega_0)\rightarrow{\cal P}_0(\omega_0)$, given by 
\begin{eqnarray}\label{response2}
{\cal P}_0(\omega_0)=\Gamma_0\sum^\infty_{m=\lceil\,\frac{\tilde{E}}{v\gamma}\rceil}\frac{1}{\gamma}\bigg(mv-\frac{\tilde{E}}{\gamma}\bigg)^2J^2_m\big(mv-\frac{\tilde{E}}{\gamma}\big),
\end{eqnarray}
where $\Gamma_0=\frac{g^2_-\rho_0}{2\hbar\,M_\ast\,R^2}$. Note this is the response for the analogue massless
scalar field, and thus no low-energy Lorentz violation can been seen when $v<f_c$. While for the latter there is a correction to the former case, 
${\cal P}(\omega_0)\rightarrow{\cal P}_0(\omega_0)+\Delta{\cal P}$, with
\begin{eqnarray}
\Delta{\cal P}=\frac{g^2_-\rho_0m_B}{2\pi\hbar^3\gamma}\int^{\zeta_+}_{\zeta_-}d\zeta\frac{\zeta}{f(\zeta)\sqrt{v^2-f^2(\zeta)}},
\end{eqnarray}
where $\zeta_-\in(0, \zeta_c)$ and $\zeta_+\in(\zeta_c, \infty)$ are unique solutions to $f(\zeta)=v$ in the respective intervals.
This correction means the detector sees a low-energy Lorentz violation when $v>f_c$, and alternatively the departure from the standard prediction of Unruh effect 
appears as a consequence of the Lorentz violation.

%%%%%%%%%%%%%%%%%%%%%%%%%%%%%%%%%%%%%%%%%%%%%%%%%%%%%%%%%%%
\emph{Experimental implementation.---} Recent experimental advances have allowed for groundbreaking observations of strongly dipolar BEC,  
its excitation spectrum displaying roton minimum, and dynamics of impurity immersed in BEC \cite{PhysRevLett.107.190401, PhysRevLett.108.210401, cite-key, roton-mode, PhysRevLett.122.183401, PhysRevLett.126.193002, PhysRevLett.123.050402, PhysRevX.11.011037, RevModPhys.91.035001, impurity1, impurity2, PhysRevLett.126.033401}.
These experiments hold promise to realize our experimental scenario proposed above.
Specifically, we can consider a single $^{87}\mathrm{Rb}$ atom immersed in a BEC of $\mathrm{Dy}$ atom \cite{PhysRevLett.107.190401, PhysRevLett.126.193002, cite-key} which possesses a magnetic dipole moment of $10\mu_B$ with $\mu_B$ being the Bohr magneton. 
The condensate density is assumed to be $\rho_0\sim4.4\times10^3 \mathrm{\mu m}^{-2}$; 
The observer trajectory radius $R\sim10 \mathrm{\mu m}$; A typical trap frequency $\omega_z=2\pi\times10^3 \mathrm{Hz}$,
the corresponding harmonic oscillator width is $d_z\simeq0.25\times10^{-6}\mathrm{m}$. Fig. \ref{fig2} displays the transition rate in \eqref{MFTR}, and 
clearly shows the spontaneous excitation of the detector as a result of the circular acceleration in a Minkowski vacuum. This would be viewed as the circular Unruh effect:  
Thermal bath is predicted for an accelerated detector moving through the inertial vacuum. 
In addition, the transition rates clearly show the deviation from the LI field case, occurring for strongly dipolar interactions. Specifically, when the field 
slightly deviates from the LI case, e.g., $A=0.1$ case, their corresponding transition rates share similar behaviors. When this deviation becomes stronger, the excitation rates increase and behave sharply differently compared with the LI case, especially in the high velocity and low energy spacing regimes. However, for the low velocity case and large energy spacing of the detector, the excitation rates change slightly in the presence of Lorentz violation (i.e., they are not so sensitive to the breaking of Lorentz symmetry), since in such cases the detector is harder to excite. 
%The low-energy Lorentz violation from high-energy modified dispersion in circular motion has been studied in Ref. \cite{PhysRevD.97.025008}, where a particular %dispersion relation has been chosen and different Lorentz-breaking scales $M_\ast$ have been assumed during analysis. Unlike that scenario, in our proposal the %Lorentz-breaking scale $M_\ast$ is assumed to be fixed while the Lorentz-breaking dispersion relation would be adjusted by controlling the ratio of dipolar and contact %interactions between atoms or molecules. How the different deviations from the LI dispersion relation responds to a fixed Lorentz-breaking scale was present here, which %allows us to qualitatively demonstrate whether the Unruh effect is universal in the presence of Lorentz-breaking physics.

\begin{figure}
\centering
\includegraphics[width=0.4\textwidth]{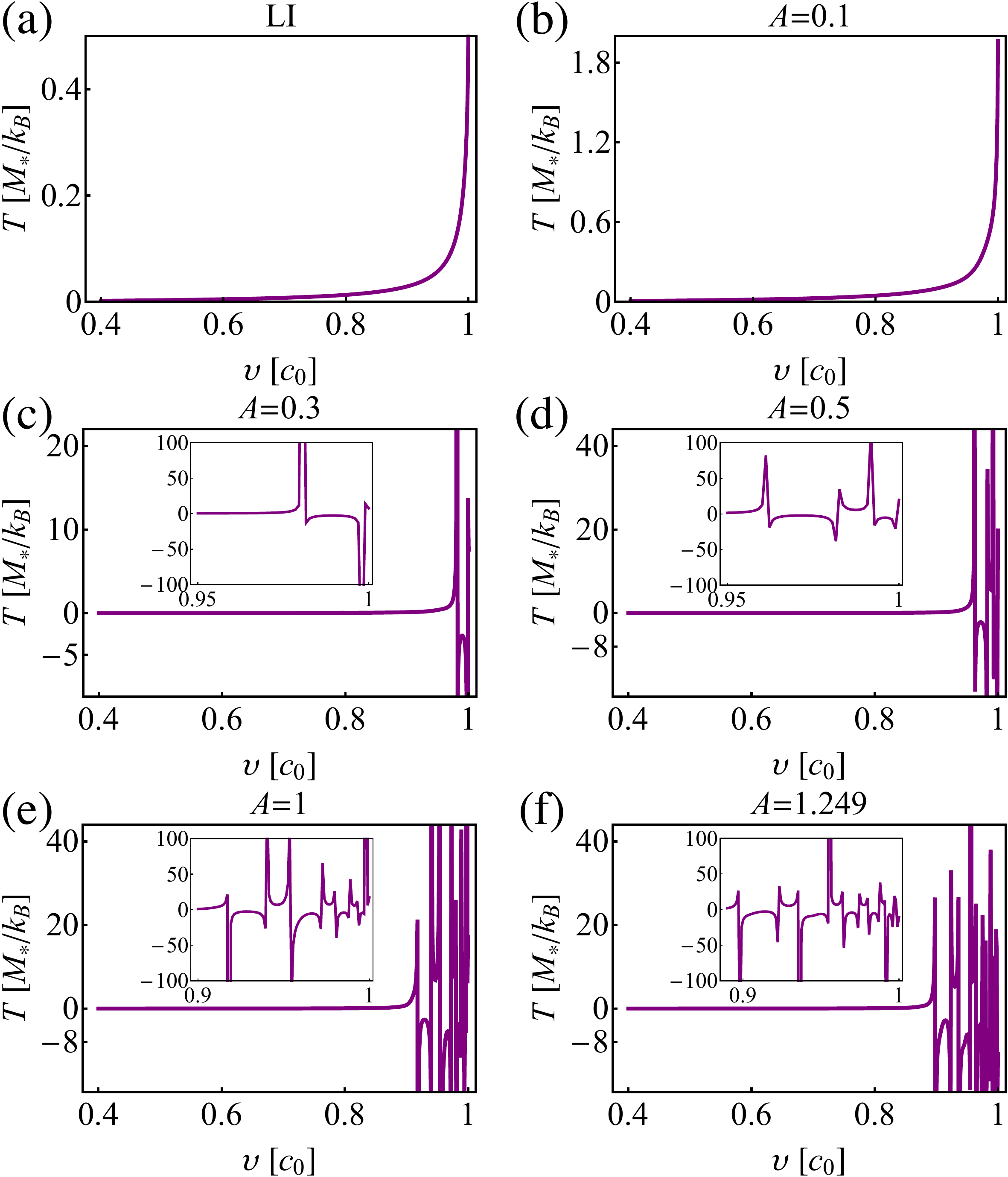}
\caption{The temperature defined in \eqref{temperature} as a function of the detector's velocity $v$ with fixed energy spacing of the detector, $\omega_0=M_\ast/\hbar$. 
Insets are the details of the corresponding oscillation parts. Here we assume that the DDI dominance case ($R=\sqrt{\pi/2}$) 
is vailid.}\label{fig3}
\end{figure}

To further check whether the Unruh effect is robust to the breaking of Lorentz symmetry, we naively define the Unruh temperature $T$ to estimate the fluctuations sampled by the impurity, using the Einstein's detailed balanced condtion,
\begin{eqnarray}\label{temperature}
T=\hbar\omega_0k^{-1}_B\ln^{-1}\bigg(\frac{{\cal P}(-\omega_0)}{{\cal P}(\omega_0)}\bigg).
\end{eqnarray}
This temperature is independent of $\omega_0$ for uniformly linearly accelerated detectors, given by $T_\text{U}=\frac{\hbar\,a}{2\pi\,k_\text{B}\,c}$, since whose response function satisfies the Kubo-Martin-Schwinger condition \cite{KMS1, KMS2, KMS3}. 
For the circular acceleration cases, such definition of $T$ may depend on $\omega_0$, however, keeps monotonous 
increase via the effective acceleration parameter for the LI case \cite{PhysRevD.102.085006}. We here plot this temperature registered by the impurity coupled to 
the analogue LVQF as shown in Fig. \ref{fig3}. For the LI case, the temperature increases monotonously with the increase of the detector's speed 
$v$ as expected. If the spectrum of the field deviates from the LI case slightly, e.g., $A=0.1$ case, the present temperature behaves similarly as the LI case 
but with a larger magnitude. Remarkably, when the spectra of the analogue LVQF strongly deviate from the LI spectrum, the temperature first increases with the increase of the detector's speed and then oscillates if the 
detector's speed exceeds a critical value which depends on the degree of the deviation. The counterintuitive oscillation phenomenon means that Lorentz violation may 
break the thermal characteristic of Unruh effect, and even cause the analogue anti-Unruh effect \cite{BRENNA2016307}: \emph{Unruh temperature decreases with the 
increase of acceleration}. Besides, the effective negative temperature \cite{PhysRev.81.279, PhysRev.103.20} occurs during the oscillation, which implies that as a result of Lorentz violation,
the corresponding detector's state is characterized by an inverted occupation distribution, where excited state is 
populated more than ground state.

%%%%%%%%%%%%%%%%%%%%%%%%%%%%%%%%%%%%%%%%%%%%%%%%%%%%%%%%%%%
\emph{Conclusions.---} We present a concrete experimental proposal to test how the Lorentz violation affects the circular Unruh effect using an impurity immersed in a 
dipolar BEC. We find that if the spectra of quantum field deviate from the LI case strongly, the transition rates and the predicted temperature of the 
analogue Unruh-DeWitt detector behave quite differently compared with the LI case, and the Lorentz violation even more may induce the counter-intuitive anti-Unruh effect on certain conditions.
Our preliminary estimates indicate that the proposed experimental implementation of the analogue circular Unruh effect and its interaction with the Lorentz-breaking physics 
is within reach of current state-of-the-art ultracold-atom experiments.

Our proposed quantum fluid platform may also allow us in the experimentally accessible regime
to explore open questions concerning Unruh effect \cite{PhysRevLett.101.110402}, and 
why its robustness to high energy modifications of the dispersion relation \cite{PhysRevLett.123.041601}
behaves differently from that of 
its equivalence principle dual---Hawking effect \cite{PhysRevD.51.2827}.
In addition, two impurities could be used as detectors to explore correlations harvest from the quantum vacuum of analogue quantum fields \cite{PhysRevD.92.064042}.

%%%%%%%%%%%%%%%%%%%%%%%%%%%%%%%%%%%%%%%%%%%%%%%%%%%%%%%%%
\begin{acknowledgments}
This work was supported by the National Key R\&D Program of China (Grant No. 2018YFA0306600),  and Anhui Initiative in Quantum Information Technologies (Grant No. AHY050000).
ZT was supported by the National Natural Science Foundation of 
China under Grant No. 11905218, and the CAS Key Laboratory for Research in Galaxies and Cosmology, Chinese Academy of Science (No. 18010203). 
\end{acknowledgments}

\bibliography{Rotating-Unruh-Detector-Dipolar-BEC}
\onecolumngrid
\vspace{1.5cm}

\newpage
%==========================================================================================================
%==========================================================================================================
%==================================== Supplemental Material ====================================================
%==========================================================================================================
%==========================================================================================================
\pagebreak
\clearpage
\widetext

\begin{center}
\textbf{\large Supplementary Material}
\end{center}

\setcounter{equation}{0}
\setcounter{section}{0}
\setcounter{page}{1}
\makeatletter
\renewcommand{\theequation}{S\arabic{equation}}

\section{Unruh effect of accelerated detector} 
We here simply review the Unruh effect corresponding to the  linear acceleration and circular motion cases.
In quantum field theory, usually quantum field is probed with a 
linearly coupled Unruh-DeWitt detector \cite{PhysRevD.14.870, birrell1984quantum}, which is described by a
localized system with internal levels $|g\rangle$ and $|e\rangle=\sigma^{+}|g\rangle$ and the energy gap $\omega_0$, 
moving along a trajectory $(t(\tau), {\bf x}(\tau))$ with $\tau$ being the detector's proper time. The detector couples with a scalar field 
$\phi$, initially in its vacuum state, through 
\begin{eqnarray}\label{interaction}
H_\text{int}=g\chi(\tau)(e^{i\omega_0\tau}\sigma^+e^{-i\omega_0\tau}\sigma_-)\phi({\bf x}(\tau), t(\tau)),
\end{eqnarray}
where $g$ is the coupling parameter and $\chi(\tau)$ is a real-valued smooth switching function that specifies how the interaction is 
turned on and off. In the first-order perturbation theory, the probability for the detector to be excited 
from its ground $|g\rangle$ to the excited state $|e\rangle$ is proportional to the response function,
\begin{eqnarray}
\nonumber
{\cal F}(\omega_0)=\frac{g^2}{\hbar^2}\int\int\,d\tau\,d\tau^\prime\chi(\tau)\chi(\tau^\prime)e^{-i\omega_0(\tau-\tau^\prime)}{\cal W}(\tau, \tau^\prime),
\end{eqnarray}
where ${\cal W}(\tau, \tau^\prime)=\langle0|\phi(t(\tau), {\bf x}(\tau))\phi(t(\tau^\prime), {\bf x}(\tau^\prime))|0\rangle$ denotes the Wightman
function evaluated along the detector's trajectory. If we consider the scenario where the detectors's trajectories and quantum field state are stationary,
in this sense that ${\cal W}(\tau, \tau^\prime)$ depends on its arguments only through the difference $s=\tau-\tau^\prime$. We
may then calculate the transition probability per unit time (or the transition rate) by dividing ${\cal F}(\omega_0)$ with
respective to the total interaction time and letting this interaction time tend to infinity, finally find
\begin{eqnarray}\label{transition-rate}
{\cal P}(\omega_0)={\cal F}_\text{rate}(\omega_0)=\frac{g^2}{\hbar^2}\int_{-\infty}^\infty\,ds\,e^{-i\omega_0s}{\cal W}(s, 0).
\end{eqnarray}
Note \eqref{transition-rate} denotes the excitation rate from the detector's ground state to its excited state.

The Wightman function for a massless scalar field in Minkowski spacetime  is analytically  ${\cal W}(\tau, \tau^\prime)=-\frac{\hbar\,c_0}{4\pi^2}\frac{1}{(c_0t(\tau)-c_0t^\prime(\tau^\prime))^2-({\bf x}(\tau)-{\bf x}^\prime(\tau^\prime))^2}$. For the linearly accelerated trajectory, $x(\tau)=\frac{c^2_0}{a}\cosh\frac{a\tau}{c_0}, t(\tau)=\frac{c_0}{a}\sinh\frac{a\tau}{c_0}$, it reduces to 
\begin{eqnarray}\label{L-Wightman}
\nonumber 
{\cal W}_l(\tau, \tau^\prime)&=&-\frac{a^2\hbar}{16\pi^2c^3_0}\bigg\{\sinh\bigg[\frac{a}{2c_0}(\tau-\tau^\prime)\bigg]\bigg\}^{-2}
\\       
&=&-\frac{\hbar}{4\pi^2c_0}(\tau-\tau^\prime)^{-2}\bigg(1+\frac{1}{12}\bigg[\frac{a}{c_0}(\tau-\tau^\prime)\bigg]^2+\frac{1}{360}\bigg[\frac{a}{c_0}(\tau-\tau^\prime)\bigg]^4+\dots\bigg)^{-1}.
\end{eqnarray}
Together with \eqref{transition-rate}, one can find that the accelerated detector becomes thermalized
with the transition probabilities satisfying 
${\cal P}_\text{excitation}/{\cal P}_\text{de-excitation}={\cal P}(\omega_0)/{\cal P}(-\omega_0)=e^{-\hbar\omega_0/k_\text{B}T_\text{U}}$, where $T_\text{U}=\frac{\hbar\,a}{2\pi\,c_0k_B}$ is the Unruh temperature. Note that observation of Unruh effect in the practice remains a challenged problem because of the extreme requirement of high linearly acceleration for typical detector's transition.

Going beyond the linear acceleration scenario, the circular motion with a constant radial acceleration could also produce an approximately thermal spectrum 
 \cite{BELL1983131, BELL1987488,UNRUH1998163}, presenting the circular Unruh effect. Specifically, if the detector moves along a circular trajectory $(c_0t(\tau), {\bf x}(\tau))=(c_0\gamma\tau, R\cos(\Omega\gamma\tau), R\sin(\Omega\gamma\tau), 0)$, with constant radius $R$, the usual relativistic factor $\gamma=(1-v^2/c^2_0)^{-1/2}=(1-\beta^2)^{-1/2}$, the angular velocity $\Omega=v/R$ and the corresponding acceleration $a=v^2\gamma^2/R=\Omega^2\gamma^2R$, one can find the corresponding Wightman function 
 \begin{eqnarray}\label{C-Wightman}
\nonumber 
{\cal W}_c(\tau, \tau^\prime)&=&-\frac{\hbar}{4\pi^2c_0}\bigg\{\gamma^2(\tau-\tau^\prime)^2-\bigg(2\frac{c_0}{a}\beta^2\gamma^2\bigg)^2\bigg(\sin\bigg[\frac{a}{2c_0}\frac{\tau-\tau^\prime}{\beta\gamma}\bigg]\bigg)^2\bigg\}^{-1}
\\       
&=&-\frac{\hbar}{4\pi^2c_0}(\tau-\tau^\prime)^{-2}\bigg(1+\frac{1}{12}\bigg[\frac{a}{c_0}(\tau-\tau^\prime)\bigg]^2-\frac{1}{360\beta^2\gamma^2}\bigg[\frac{a}{c_0}(\tau-\tau^\prime)\bigg]^4+\dots\bigg)^{-1}.
\end{eqnarray}
Comparing the Wightman function \eqref{C-Wightman} with \eqref{L-Wightman}, there is distinct difference: an extra parameter, $\beta$ or $\gamma$, appears in the circular motion case. This is because for circular motion the radius of the circle can be varied independently of the acceleration. Unlike the linear accelerated case, the Fourier transformation of \eqref{C-Wightman} is not analytical. However,  it, in the ultrarelativistic limit, $\gamma\gg1$, can be simplified.  Then the corresponding equilibrium transition probabilities
yields \cite{BELL1983131}
 \begin{eqnarray}
\frac{ {\cal P}_\text{excitation}}{{\cal P}_\text{de-excitation}}\approx\frac{a}{4\sqrt{3}\omega_0c_0}e^{-2\sqrt{3}\frac{\omega_0c_0}{a}}
 \end{eqnarray}
 for the $a/c_0\omega_0\ll1$ case. It leads to an effective temperature $T_\text{eff}=\frac{\hbar\,a}{2\sqrt{3}k_Bc_0}$, which is higher by a factor $\pi/\sqrt{3}$ than 
 the Unruh temperature for the linear acceleration.

Note that compared with the linear acceleration case, circular motion allows the accelerating system to remain within a finite-size laboratory for 
an arbitrarily long interaction time. Furthermore, unlike what happens in uniform linear acceleration, the proper and coordinate time for the circular motion are related by a time-independent gamma faction, which will be crucial when estimating the experimental feasibility for detecting the analogue circular Unruh effect.

\section{Derivation of the transition rate}
The density fluctuations of dipolar BEC is of the form,
\begin{eqnarray}
\delta\hat{\rho}(t,\br)=\sqrt{\rho_0}\int\frac{d^2\bk}{(2\pi)^2}(u_\bk+v_\bk)\big[\hat{b}_\bk(t)e^{i\bk\cdot\br}+\hat{b}^\dagger_\bk(t)e^{-i\bk\cdot\br}\big].
\end{eqnarray}
We can use it to calculate the analogue correlation function of the density fluctuations
\begin{eqnarray}\label{final-CF}
\nonumber
\langle0|\delta\hat{\rho}(t,\br)\delta\hat{\rho}(t^\prime,\br^\prime)|0\rangle&=&\frac{\rho_0}{(2\pi)^4}\int\int\,d^2\bk\,d^2\bk^\prime(u_{k}+v_{k})(u_{k^\prime}+v_{k^\prime})
\langle0|\big[\hat{b}_\bk(t)e^{i\bk\cdot\br}+\mathrm{h.c.}\big]\big[\hat{b}_{\bk^\prime}(t^\prime)e^{i\bk^\prime\cdot\br^\prime}+\mathrm{h.c.}\big]|0\rangle
\\  \nonumber 
&=&\frac{\rho_0}{(2\pi)^2}\int\int\,d^2\bk\,d^2\bk^\prime(u_{k}+v_{k})(u_{k^\prime}+v_{k^\prime})e^{-i\omega_\bk\,t+i\omega_{\bk^\prime}t^\prime}e^{i\bk\cdot\br-i\bk^\prime\cdot\br^\prime}\delta^2(\bk-\bk^\prime)
\\  \nonumber 
&=&\frac{\rho_0}{(2\pi)^2}\int\,d^2\bk(u_{k}+v_{k})^2e^{-i\omega_\bk(t-t^\prime)}e^{i\bk\cdot(\br-\br^\prime)}
\\  \nonumber 
&=&\frac{\rho_0}{(2\pi)^2}\int^\infty_0dk\int^{2\pi}_0d\theta\,k(u_{k}+v_{k})^2e^{-i\omega_\bk(t-t^\prime)}e^{i|\bk||\br-\br^\prime|\cos\theta}
\\  
&=&\frac{\rho_0}{(2\pi)}\int^\infty_0dk\,k(u_{k}+v_{k})^2e^{-i\omega_\bk(t-t^\prime)}J_0\big(|\bk||\br-\br^\prime|\big),
\end{eqnarray} 
where $J_0(x)$ is the Bessel function of the first kind. For the uniform circular motion case, the trajectory is 
$(t(\tau), \br(\tau))=(\gamma\tau, R\cos(\gamma\Omega\tau), R\sin(\gamma\Omega\tau))$, and it is easy to find $t-t^\prime=\gamma(\tau-\tau^\prime)=\gamma\,s$ and $|\br-\br^\prime|=2R\sin\frac{\gamma\Omega}{2}s$. Inserting this trajectory into
the above correlation function of the density fluctuations, we can calculate the transition rate 
\begin{eqnarray}\label{FTR}
 \nonumber
{\cal P}(\omega_0)&=&\frac{g^2_-\rho_0}{2\pi\hbar^2}
\int\,ds\int^\infty_0dk\,k(u_k+v_k)^2J_0\bigg(2Rk\sin\frac{\gamma\Omega}{2}s\bigg)e^{-i(\omega_0+\gamma\omega_\bk)s}
\\    \nonumber
&=&\frac{g^2_-\rho_0}{2\pi\hbar^2}
\int\,ds\int^\infty_0dk\,k(u_k+v_k)^2\sum^{\infty}_{m=-\infty}J^2_m(Rk)e^{im\gamma\Omega\,s}e^{-i(\omega_0+\gamma\omega_\bk)s}
\\    \nonumber
&=&\frac{g^2_-\rho_0}{\hbar^2\gamma}\int^\infty_0dk\,k(u_k+v_k)^2
\sum^{\infty}_{m=-\infty}J^2_m(Rk)\delta\big(\frac{\omega_0}{\gamma}+\omega_\bk-m\Omega\big)
\\ 
&=&\frac{g^2_-\rho_0m_B}{2\hbar^3}\int^\infty_0d\zeta\frac{\zeta^2}{\gamma\,f(\zeta)}\sum^{\infty}_{m=-\infty}J^2_m(\tilde{M}\zeta)
\delta\bigg(\zeta\,f(\zeta)-\frac{1}{\tilde{M}}\bigg(mv-\frac{\tilde{E}}{\gamma}\bigg)\bigg),
\end{eqnarray}
where $\zeta=\hbar\,c_0k/M_\ast$, $\tilde{M}=RM_\ast/\hbar\,c_0$, $\tilde{E}=R\omega_0/c_0$, and $v=R\Omega/c_0$ are dimensionless parameters. Note that to 
derive the second equality the identity, $J_0(2a\sin\,x)=\sum_{m\in\mathbf{Z}}J^2_m(a)e^{2imx}$, has been used.

\subsection{Thermal circular Unruh effect}
If $R_0=0$, it means only contact interaction happens between atoms or molecules in Bose gas, we can find the excitation spectrum has the form, $\omega_\bk=c_0k\sqrt{1+\zeta^2/4}$, which is ``relativistic" \cite{PhysRevLett.101.110402}. $\omega_\bk\approx\,c_0k$, for $k\ll\,k_c=M_\ast/c_0\hbar$. Then, \eqref{FTR}
reduces to,
\begin{eqnarray}\label{massless-case}
\nonumber
{\cal P}(\omega_0)&=&\frac{g^2_-\rho_0}{2\pi\hbar^2}
\int\,ds\int^\infty_0dk\,k(u_k+v_k)^2J_0\bigg(2Rk\sin\frac{\gamma\Omega}{2}s\bigg)e^{-i(\omega_0+\gamma\omega_\bk)s}
\\    \nonumber
&=&\frac{g^2_-\rho_0}{2\pi\hbar^2}
\int\,ds\int^\infty_0dk\,k\frac{\hbar^2k^2/2m_B}{\hbar\,c_0k}J_0\bigg(2Rk\sin\frac{\gamma\Omega}{2}s\bigg)e^{-i\gamma\,c_0ks}e^{-i\omega_0s}
\\    \nonumber
&=&\frac{g^2_-\rho_0}{4\pi\hbar\,m_Bc_0}
\int\,ds\int^\infty_0dk\,k^2J_0\bigg(2Rk\sin\frac{\gamma\Omega}{2}s\bigg)e^{-i\gamma\,c_0ks}e^{-i\omega_0s}
\\    
&=&\frac{g^2_-\rho_0}{4\pi\hbar\,m_Bc_0}
\int\,ds\frac{-2\gamma^2c^2_0s^2-4R^2\sin^2\frac{\gamma\Omega}{2}s}{\big(-\gamma^2c^2_0s^2+4R^2\sin^2\frac{\gamma\Omega}{2}s\big)^{5/2}}e^{-i\omega_0s}.
\end{eqnarray}
In the ultrarelativistic limit, $\gamma\gg1$, we can do the same process as in \cite{BELL1983131} to further calculate the above transition rate, which is given by
\begin{eqnarray}
\nonumber
{\cal P}(\omega_0)&=&\frac{g^2_-\rho_0}{4\pi\hbar\,m_Bc_0}\int\,ds\frac{24i(-1+3\gamma^2)}{24c^3_0s^3+5a^2c_0s^5}e^{-i\omega_0s}
\\  
&=&\frac{5g^2_-\rho_0}{96\sqrt{2\pi}\hbar\,m_Bc^6_0}(3\gamma^2-1)a^2\bigg[1-\exp\big(-\sqrt{\frac{24}{5}}\frac{c_0\omega_0}{a}\big)+\frac{12c^2_0\omega^2_0}{5a^2}\bigg].
\end{eqnarray}
Furthermore, assuming the energy splitting of the detector to be not too small $\omega_0\gg\,a/c_0$, we can find the equilibrium population of the upper level 
relative to the lower is 
\begin{eqnarray}
\frac{{\cal P}(\omega_0)}{{\cal P}(-\omega_0)}=\frac{12c^2_0\omega^2_0}{5a^2}\exp\big(-\sqrt{\frac{24}{5}}\frac{c_0\omega_0}{a}\big),
\end{eqnarray} 
leading to an effective temperature 
\begin{eqnarray}
T_\text{eff}=\frac{\sqrt{5}\hbar\,a}{2\sqrt{6}k_Bc_0}.
\end{eqnarray}

\subsection{Correction to the LI case}

As shown above, the transition rate of the detector from its ground state to excited state is
\begin{eqnarray}\label{final-TR1}
{\cal P}(\omega_0)=\frac{g^2_-\rho_0m_B}{2\hbar^3}\int^\infty_0d\zeta\frac{\zeta^2}{\gamma\,f(\zeta)}\sum^{\infty}_{m=-\infty}J^2_m(\tilde{M}\zeta)\delta\bigg(\zeta\,f(\zeta)-\frac{1}{\tilde{M}}\bigg(mv-\frac{\tilde{E}}{\gamma}\bigg)\bigg),
\end{eqnarray}
where $\tilde{M}=RM_\ast/\hbar\,c_0$, $\tilde{E}=R\omega_0/c_0$, and $v=R\Omega/c_0$ are dimensionless parameters.

We consider the stable dipolar BEC and thus $f(\zeta)$ is smooth and strictly positive, and $f(\zeta)\rightarrow1$ as $\zeta\rightarrow0$. 
We also consider the scenario where the only stationary point of $f$ is a global minimum at $g=g_c>0$, written as $f_c=f(g_c)$ with
$0<f_c<1$. In such case, $f^\prime(\zeta)<0$ for $0<\zeta<\zeta_c$ and $f^\prime(\zeta)>0$ for $\zeta>\zeta_c$.
In addition to that, we also assume that $\zeta\,f(\zeta)$ is a monotonely increasing function of $\zeta$, which means 
$(\zeta\,f(\zeta))^\prime=\zeta\,f^\prime(\zeta)+f(\zeta)>0$ for $\zeta>0$. Then we can 
perform the integral in Eq. \eqref{final-TR1}, and find 
\begin{eqnarray}\label{response1}
{\cal P}(\omega_0)=\frac{g^2_-\rho_0m_B}{2\hbar^3\gamma}\sum^\infty_{m=\lceil\,\frac{\tilde{E}}{v\gamma}\rceil}\frac{\zeta^2_m/f(\zeta_m)}{(\zeta\,f(\zeta))^\prime\vert_{\zeta=\zeta_m}}J^2_m(\tilde{M}\zeta_m),
\end{eqnarray}
where $\zeta_m$ is the unique solution to $\zeta\,f(\zeta)=\frac{1}{\tilde{M}}\big(mv-\frac{\tilde{E}}{\gamma}\big)$.
In the limit $\tilde{M}\rightarrow\infty$, we find that the limit is qualitatively different for $0<v<f_c$ and $f_c<v<1$, as found in Refs. \cite{PhysRevD.97.025008}:
\begin{eqnarray}\label{response2}
{\cal P}_0(\omega_0)=\frac{g^2_-\rho_0}{2\hbar\,M_\ast\,R^2\gamma}\sum^\infty_{m=\lceil\,\frac{\tilde{E}}{v\gamma}\rceil}\bigg(mv-\frac{\tilde{E}}{\gamma}\bigg)^2J^2_m\big(mv-\frac{\tilde{E}}{\gamma}\big),
\end{eqnarray}
for the $0<v<f_c$ case. Note that in this case the detector sees no low-energy Lorentz violation: the corresponding response is the same as that for the usual massless
scalar field. For $f_c<v<1$, ${\cal P}(\omega_0)\rightarrow{\cal P}_0(\omega_0)+\Delta{\cal P}$ as $\tilde{M}\rightarrow\infty$, where 
\begin{eqnarray}
\Delta{\cal P}=\frac{g^2_-\rho_0m_B}{2\pi\hbar^3\gamma}\int^{\zeta_+}_{\zeta_-}d\zeta\frac{\zeta}{f(\zeta)\sqrt{v^2-f^2(\zeta)}},
\end{eqnarray}
and $\zeta_-\in(0, g_c)$ and $\zeta_+\in(g_c, \infty)$ are unique solutions to $f(\zeta)=v$ in the respective intervals.

Note that when $v>f_c$ and $\tilde{M}$ is large, the Lorentz-breaking contribution to the sum in Eq. \eqref{response1} comes from 
values of $m$ that are comparable to $\tilde{M}$. In conclusion, Lorentz violation of quantum fields would affect the transition rate 
of the Unruh-DeWitt detector: The detector in circular motion in the preferred inertial frame sees
a large low-energy Lorentz violation when its orbital speed exceeds the critical value $f_c$.

\end{document}